\documentclass[a4paper,12pt]{article}
\usepackage{graphicx}
\usepackage{authblk}
\usepackage{amsmath}
\usepackage{multicol}
\usepackage{multirow}
\usepackage{setspace}
\usepackage{array}
\usepackage{float}
\usepackage[top=2.5cm,bottom=2.5cm,left=1.5cm,right=1.5cm]{geometry}
\begin{document}

\title{Self-regulating genes. Exact steady state solution by using Poisson Representation.}
\author[1]{Istv\'an P. Sug\'ar}
\author[2]{Istv\'an Simon}
\affil[1]{\small Department of Neurology and Center for Translational Systems Biology, Ichan School of Medicine at Mount Sinai, New York, NY 10029}
\affil[2]{\small Institute of Enzymology, Research Center for Natural Sciences, Hungarian Academy of Sciences, Budapest, Hungary}
\date{\the\year}
\maketitle
{e-mail: istvan.sugar@mssm.edu, simon@enzim.hu}

\section*{Abstract}

Systems biology studies the structure and behavior of complex gene regulatory networks. One of its aims is to develop a quantitative understanding of the modular components that constitute such networks. The self-regulating gene is a type of auto regulatory genetic modules which appears in over 40\% of known transcription factors in E. coli. In this work, using the technique of Poisson Representation, we are able to provide exact steady state solutions for this feedback model. By using the methods of synthetic biology (P.E.M. Purnick and Weiss, R., Nature Reviews, Molecular Cell Biology, 2009, 10: 410-422) one can build the system itself from modules like this.

\section{Introduction}

An important aim of systems biology is to dissect gene regulatory networks to modular components [1]. In the last decade, by using graph theoretical methods, important functional modules were found [2-4]. These extracted, relatively simple modules can be studied theoretically. Once the modules are well characterized they can be coupled to increase network complexity and to better understand cellular behavior.

Because of the low reactant number in the cell the fluctuations should be included into the theoretical description of a module, i.e. a stochastic theory is needed. The stochastic processes can be either simulated by using Gillespie’s[5] method or by analytically solving the respective master equations. 

Among the identified modules, feedback loops are present in many cellular networks [2, 4, 6-7] and self-regulating genes are simple examples of these modules. So far two models of self-regulating genes have been developed and studied theoretically: Hornos’ model [8] and Iyer-Biswas’ model [9]. These are two-state models of a gene which produces a protein that regulates its own activity. In Iyer-Biswas’ model the gene is either in ‘off ‘state (no protein production at all) or in ‘on’ state (proteins are produced). In the positive (negative) feedback model, the amount of protein produced proportionally increases the propensity of the gene to be in the ‘on (off)’ states.

In Hornos’ model the gene is in $\alpha$  state when the operator site is free, while it is in $\beta$  state when the protein is bound to the operator site. The rates for protein production,$g_\alpha$ and $g_\beta$, are different for the free and bound state. $g_\alpha > g_\beta$ and $g_\alpha < g_\beta$ in the case of a self-repressing and self-activating gene, respectively.

In this article we consider a slightly modified version of the self-regulating gene model originally developed by Hornos et al. [8]. We also correct the master equations of Hornos’ model and calculate the steady state solution by using the method of Poisson Representation[10-11]. Note that the same corrections of the master equations have been made by Grima et al.[12],  and solved the equations by using the generating function method.

\section{Model}

In our self-regulating gene model a single gene produces a protein that represses or enhances its own activity. The six elementary reactions where $\alpha$ and $\beta$ denote the protein unbound and protein bound states of the gene, respectively are:

\begin{eqnarray}
\alpha+P  & \xrightarrow{\hspace{1em}h\hspace{1em}}        &   \beta      \text{\hspace{4em}protein binding}\\ 
\beta     &  \xrightarrow{\hspace{1em}f\hspace{1em}}       &   \alpha+P   \text{\hspace{2em}protein dissociation}\\
\beta     & \xrightarrow{\hspace{1em}k_b\hspace{1em}}      &   \alpha     \text{\hspace{4em}degradation of bound protein}\\
\alpha    & \xrightarrow{\hspace{1em}g_\alpha\hspace{1em}} &   \alpha+P   \text{\hspace{2em}unbound gene produces protein}\\
\beta     & \xrightarrow{\hspace{1em}g_\beta\hspace{1em}}  &   \beta+P    \text{\hspace{2em}bound gene produces protein}\\
P         & \xrightarrow{\hspace{1em}k\hspace{1em}}        &   \emptyset  \text{\hspace{4em}degradation of unbound protein} 
\end{eqnarray}

where $P$ is the unbound protein, $h$ is the bimolecular rate of protein binding to the gene, $f$ is the unimolecular rate of protein release from the gene. The rates for protein production $g_\alpha$ and $g_\beta$ are different for protein free and protein bound states of the gene. $g_\alpha > g_\beta$ and $g_\alpha < g_\beta$ in the case of self-repression and self-activation, respectively. $k$ is the degradation rate of an unbound protein.

The above reactions and notations are similar to the ones in Hornos’ model [8] except Eq.3 describes the degradation of a bound protein with a rate of $k_b$. In Hornos’ model this reaction is taken into consideration only when there is no unbound protein present and the protein degradation rate is $k_b=k$. In all subsequent version of Hornos’ model [13-16] the degradation rate of the bound protein is taken to be zero. In our model, for the sake of generality, any value of $k_b$ is acceptable. This is the case because the degradation rate may depend on the actual gene-protein interaction.

\subsection{Master equations}

The probability of finding $n$ proteins (unbound) in the system at time $t$ can be calculated by solving the master equations. The master equations of our model at $n \ge 0$ are:

\begin{eqnarray}
\frac{dP_\alpha(n,t)}{dt} & = & f P_\beta(n-1,t)+ k_b P_\beta(n,t) + g_\alpha P_\alpha(n-1,t) \notag \\
& & + k[n+1] P_\alpha(n+1,t) - [ hn + g_\alpha + kn ] P_\alpha(n,t) \\ \notag \\
\frac{dP_\beta(n,t)}{dt} & = & k[n+1 ]P_\beta(n+1,t) + g_\beta P_\beta(n-1,t) \notag \\
& & + h[n+1] P_\alpha(n+1,t) - [ f + k_b + g_\beta + nk ] P_\beta(n,t)
\end{eqnarray}

where $P_\alpha(n,t)$ and $P_\beta(n,t)$ are the individual probabilities that the gene is unbound and bound, respectively, while immersed in a solution containing $n$ unbound proteins at time $t$. In addition it is physically plausible to make the following additional restrictions:

\begin{equation} 
P_\alpha(n,t)=P_\beta(n,t)\equiv0 \text{\hspace{1em}at\hspace{1em}}n<0
\end{equation}

Since we intend to get the steady state solutions of the above master equations, initial conditions are not needed.

In the Supplemental Material (Part 1) detailed comparisons are made between our master equations, Eqs.7-8, and the master equations utilized in the works of Hornos and his coworkers [8, 13-16]. We point out differences between the two sets of equations that are either because of the extra reaction in our model (Eq.3) or because of the erroneous terms in Hornos’ equations.

\subsection{Equations of generating functions}

In order to get the steady state solutions of the master equations, we rewrite Eqs.7-9 in the form of generating equations, where the generating functions are defined as:

\begin{equation}
G_i(s,t) \equiv \sum\limits_{n=0}^{\infty} s^n P_i(n,t)
\end{equation}
where $i = \alpha$ or $\beta$.

We obtain the following equation for generating function $G_\alpha$:
\begin{eqnarray}
\frac{\partial G_\alpha(s,t)}{\partial t} & = & fsG_\beta(s,t) + k_b G_\beta(s,t) + g_\alpha sG_\alpha(s,t) \notag\\
& & + k \frac{\partial{G_\alpha(s,t)}}{\partial s} - (h+k) s \frac{\partial G_\alpha(s,t)}{\partial s}-g_\alpha G_\alpha(s,t)
\end{eqnarray}

and for generating function $G_\beta$:  

\begin{equation}
\frac{\partial G_\beta(s,t)}{\partial t} = k \frac{\partial G_\beta(s,t)}{\partial s} + g_\beta s G_\beta(s,t) + h \frac{\partial G_\alpha(s,t)}{\partial s} - ( f + k_b + g_\beta ) G_\beta(s,t)- k s \frac{\partial G_\beta(s,t)}{\partial s}
\end{equation}

\subsection{Poisson representation at steady state}

Let us solve Eqs.7,8 at steady state by using the method of Poisson representation [10]. The method assumes the existence of  $\rho_\alpha(\lambda)$ and $\rho_\beta(\lambda)$ functions that yield the steady state solutions, $P_\alpha(n,\infty)$ and $P_\beta(n,\infty)$, by

\begin{equation}
P_i(n,\infty) = \int\limits_A^B  d\lambda\rho_i(\lambda) e^{-\lambda} \frac{\lambda^n}{n!}
\end{equation}
where $i=\alpha$ or $\beta$.

After substituting the above forms of the probability functions into the generating functions in Eqs.11,12, we get the following equations for the $\rho_\alpha(\lambda)$ and $\rho_\beta(\lambda)$ functions:

\begin{eqnarray}
0 & = & \bigg[ f\rho_\beta(B) + \Big\{ g_\alpha - hB - kB \Big\} \rho_\alpha(B) \bigg] e^{B(s-1)} \notag\\
& & \hspace{-0.8em}- \bigg[ f\rho_\beta(A) + \Big\{g_\alpha-hA-kA\Big\}\rho_\alpha(A) \bigg] e^{A(s-1)} \notag\\
& & \hspace{-0.8em}+ \int\limits_A^B d\lambda \left\{ f\rho_\beta - f \frac{d\rho_\beta}{d\lambda}+k_b\rho_\beta -g_\alpha \frac{d\rho_\alpha}{d\lambda}+k \frac{d(\lambda\rho_\alpha)}{d\lambda} + h\frac{d(\lambda\rho_\alpha)}{d\lambda} - h\lambda\rho_\alpha \right\} e^{\lambda(s-1)} \\ \notag \\
0 & = & \Big[ g_\beta-kB \Big]\rho_\beta(B) e^{B(s-1)} - \Big[g_\beta-kA\Big]\rho_\beta(A) e^{A(s-1)} \notag \\
& & +\int\limits_A^B d\lambda \Big\{ k \frac{d(\lambda\rho_\beta)}{d\lambda} - g_\beta \frac{d\rho_\beta}{d\lambda} + h\lambda\rho_\alpha - f\rho_\beta-k_b\rho_\beta \Big\} e^{\lambda(s-1)}
\end{eqnarray}

The detailed derivation and the solutions of Eqs.14,15 are given in the Supplemental Material (Part 2).

\section{Results}

The steady state solution of the master equations, Eqs.7-8, is based on an expansion of the probability distribution in Poisson distributions (see Eq.13). The functions of the expansion coefficients, $\rho_\alpha(\lambda)$ and $\rho_\beta(\lambda)$,  are the solutions of Eqs.14,15. With the proper selection of the integration boundaries, $A$ and $B$, the first two terms of each equation, Eqs.14,15, become zero (see derivation in the Supplemental Material (Part 2)). The functions of the expansion coefficients, $\rho_\alpha(\lambda)$ and $\rho_\beta(\lambda)$ are:

\begin{eqnarray}
\rho_\alpha(\lambda ) & = & D \bigg[\frac{f+g_\beta-k\lambda}{k\lambda+h\lambda-g_\alpha}\bigg] \left| \lambda-\frac{g_\beta}{k} \right|^{F-L} \left| \lambda-\frac{g_\alpha}{k+h} \right|^G e^{E\lambda} \notag\\
& = & \bigg[ \frac{D}{k+h}\bigg] \bigg[\frac{f+g_\beta-k\lambda}{\lambda-\frac{g_\alpha}{k+h}}\bigg]  \left| \lambda-\frac{g_\beta}{k} \right|^{F-L} \left| \lambda-\frac{g_\alpha}{k+h} \right|^G e^{E\lambda} \\ \notag \\
\rho_\beta(\lambda) & = & D \left| \lambda-\frac{g_\beta}{k} \right|^{F-L}  \left|\lambda-\frac{g_\alpha}{k+h} \right|^G e^{E\lambda}
\end{eqnarray}

where $D$ is the integration constant and
\begin{eqnarray*}
E & = & \frac{h}{(k+h)}\\
F & = & \frac{f h g_\beta}{k^2 (k+h) \left[\frac{g_\alpha}{k+h} - \frac{g_\beta}{k}\right]}\\
G & = & \frac{h}{k+h} \left[ \frac{g_\alpha}{k+h} - \frac{f+g_\beta}{k} \right] \frac{\frac{g_\alpha}{k+h}}{\frac{g_\alpha}{k+h}-\frac{g_\beta}{k}}\\
L & = & \frac{(k-k_b-f)}{k}
\end{eqnarray*}

By means of a normalization condition we can calculate the integration constant, $D$ (Supplemental Material (Part 3)) 

Table 1. lists the integration boundaries, $A$ and $B$, at which the first two terms in both Eq.14 and Eq.15 become zero.

\begin{table}[H]
	\centering
	\caption{Integration Boundaries}
	\begin{tabular}[c]{|c|p{2.3cm}|c|c|p{4.3cm}|}
	\hline
	\multicolumn{2}{|c|}{\textbf{Conditions}} & \multicolumn{2}{c|}{\textbf{Integration boundaries}} & \textbf{Comments} \\
	\hline
	\multirow{2}{*}{$\frac{g_\beta}{k} < \frac{g_\alpha}{k+h}$} & \vspace{-0.3em}$F-L>0$ & \multirow{2}{*}{$A=\frac{g_\beta}{k}$} & \multirow{2}{*}{$B=\frac{g_\alpha}{k+h}$} & $\rho_\alpha \left\{\begin{array}{l l}
<0 \text{\hspace{1em}if\hspace{1em}} \frac{f+g_\beta}{k} > \lambda \\
>0 \text{\hspace{1em}if\hspace{1em}} \frac{f+g_\beta}{k}<\lambda
\end{array} \right.$ \\
	& \multirow{2}{*}{and} & & &$\rho_\beta>0$ \\
	\cline{1-1}
	\cline{3-5}
	\multirow{2}{*}{$\frac{g_\beta}{k} > \frac{g_\alpha}{k+h}$} & & \multirow{2}{*}{$A = \frac{g_\alpha}{k+h}$} & \multirow{2}{*}{$B=\frac{g_\beta}{k}$} & $\rho_\alpha \left\{\begin{array}{l l}
>0 \text{\hspace{1em}if\hspace{1em}} \frac{f+g_\beta}{k} > \lambda \\
<0 \text{\hspace{1em}if\hspace{1em}} \frac{f+g_\beta}{k}<\lambda
\end{array} \right.$ \\
	& \vspace{-2.6em}$G-1>0$ & & &$\rho_\beta>0$ \\
	\hline
	\multirow{2}{*}[-3em]{$F-L+G<0$} & $F-L>0$ \newline and \newline $G-1<0$ & \multirow{3}{*}{$A=\frac{g_\beta}{k}$} & \multirow{3}{*}{$B=\infty$} & \vspace{-0.3em}Self contradicting conditions* \\
	\cline{2-5}
		  & $F-L<0$ \newline and \newline  $G-1>0$ & \multirow{3}{*}{$A=\frac{g_\alpha}{k+h}$} & \multirow{3}{*}{$B=\infty$} & \vspace{-0.3em}Self contradicting conditions**\\
	\hline
	\multicolumn{2}{|c|}{All other conditions} & \multicolumn{2}{m{5cm}|}{Poisson representation does not exist} & \\
	\hline
	\end{tabular}
\end{table}

*From conditions $F-L+G<0$ and $G-1<0$ follows $F-L<-1$, and this contradicts with condition $F-L>0$. 

** From conditions $F-L+G<0$ and $F-L<0$ follows $G<0$, and this contradicts with condition $G-1>0$.

\subsection{Steady state solutions}

After substituting Eqs.16, 17 into Eq.13 one can calculate the steady state solutions of the master equations (Eqs.7,8). The integration boundaries in Eq.13 are given in Table 1. The integrals can be obtained in closed forms when either $\frac{g_\alpha}{k} > \frac{g_\beta}{k}=0$ or $\frac{g_\beta}{k} > \frac{g_\alpha}{k}=0$ and thus one can get $P_\alpha(n,\infty)$ and $P_\beta(n,\infty)$ in closed form too (Supplemental Material(Part 4)).

In general the integral in Eq.13 has been calculated numerically by Romberg’s method[17].  In Supplemental Material (Part 5) we point out that the solution obtained from the numerical integration is consistent with the master equations (Eqs.7,8).

The subfigures of Figure 1 show the calculated steady state probability distributions at twelve different parameter sets. All but one of these parameter sets are characteristic to self-restricting genes, i.e. where $g_\alpha > g_\beta$.

Subfigures 1a-d belong to increasing values of the $\frac{g_\beta}{k}$ parameter from 0 to 200, while the other parameters are fixed. With increasing $\frac{g_\beta}{k}$ one expects an increasing number of proteins produced by the bound gene, while with increasing number of protein the overall probability of bound gene state increases too. These qualitative expectations are supported by the calculated $P_\beta$ values (see figure legends) and the probability distributions in Subfigures 1a-d.

Subfigures 1e-h belong to increasing values of the $\frac{f}{k}$ parameter from 0.2 to 20, while the other parameters are fixed. With increasing $\frac{f}{k}$ one expects an increasing probability of the unbound gene state. Since $g_\alpha >\frac{g_\beta}{k}=0$, the increasing probability of unbound gene state results in an increasing number of proteins produced by the unbound gene. Again these qualitative expectations are supported by the calculated overall probabilities of the unbound gene state, $P_\alpha (=1-P_\beta)$, and probability distributions in Subfigures 1e-h.

Finally, subfigures 1i-l belong to increasing values of the $\frac{h}{k}$ parameter from 0.015 to 0.9, while the other parameters are fixed. With increasing $\frac{h}{k}$ one expects an increasing overall probability of the bound gene state. Since $g_\alpha > \frac{g_\beta}{k}=0$, the increasing probability of the bound gene state results in a decreasing number of proteins produced by the unbound gene. These qualitative expectations are supported by the calculated overall probabilities of the bound gene, $P_\beta$, and probability distributions in Subfigures 1i-l.

The subfigures of Figure 2 show the calculated steady state probability distributions at twelve different parameter sets. All these parameter sets are characteristic to self–activating genes, i.e. where $g_\alpha < g_\beta$. We have similar qualitative expectations regarding the change of the probability distributions with the change of parameters as in the case of the subfigures of Figure 1. However, there is one important difference between the parameter sets listed in Figure 1 and Figure 2. In Figure 2 (except for subfigures 2b-d ) both the $\frac{g_\alpha}{k}$ and $\frac{g_\beta}{k}$ parameters are small. Consequently one can expect smaller average protein numbers, $\langle n \rangle = \sum\limits_{n=0}^\infty n [ P_\alpha(n,\infty) + P_\beta(n,\infty)]$. This expectation is supported by the calculated probability distributions. In the subfigures of Figure 2 (except for subfigures 2b-d ) the average protein number is about 5, while in the case of subfigures to Figure 1 it is, in most of the cases, much higher. Finally, it is important to note that at model parameters satisfying the conditions in Table 1 the calculated probability distributions have been always unimodal ones.

\section{Conclusions}

Systems biology studies the structure and behavior of complex gene regulatory networks. One of its aims is to develop a quantitative understanding of the modular components that constitute such networks. The self-regulating gene is a type of auto regulatory genetic modules which appears in over 40\% of known transcription factors in E. coli. In this work using the technique of Poisson Representation, we are able to provide exact steady state solutions for this feedback model. The model is an extended and corrected version of  Hornos’ model [8, 13-16]. The behavior of the obtained probability distributions has been investigated at many parameter sets for both self-repressing and self-activating genes.

\section*{Acknowledgements}

I.P.S. acknowledges the support by contract NIH/NIAID HHSN272201000054C. I.S. acknowledges the support from the Hungarian Science Research Fund (OTKA-NK 100482).

\newpage

\begin{figure}
	\centering
	\includegraphics[width=15cm]{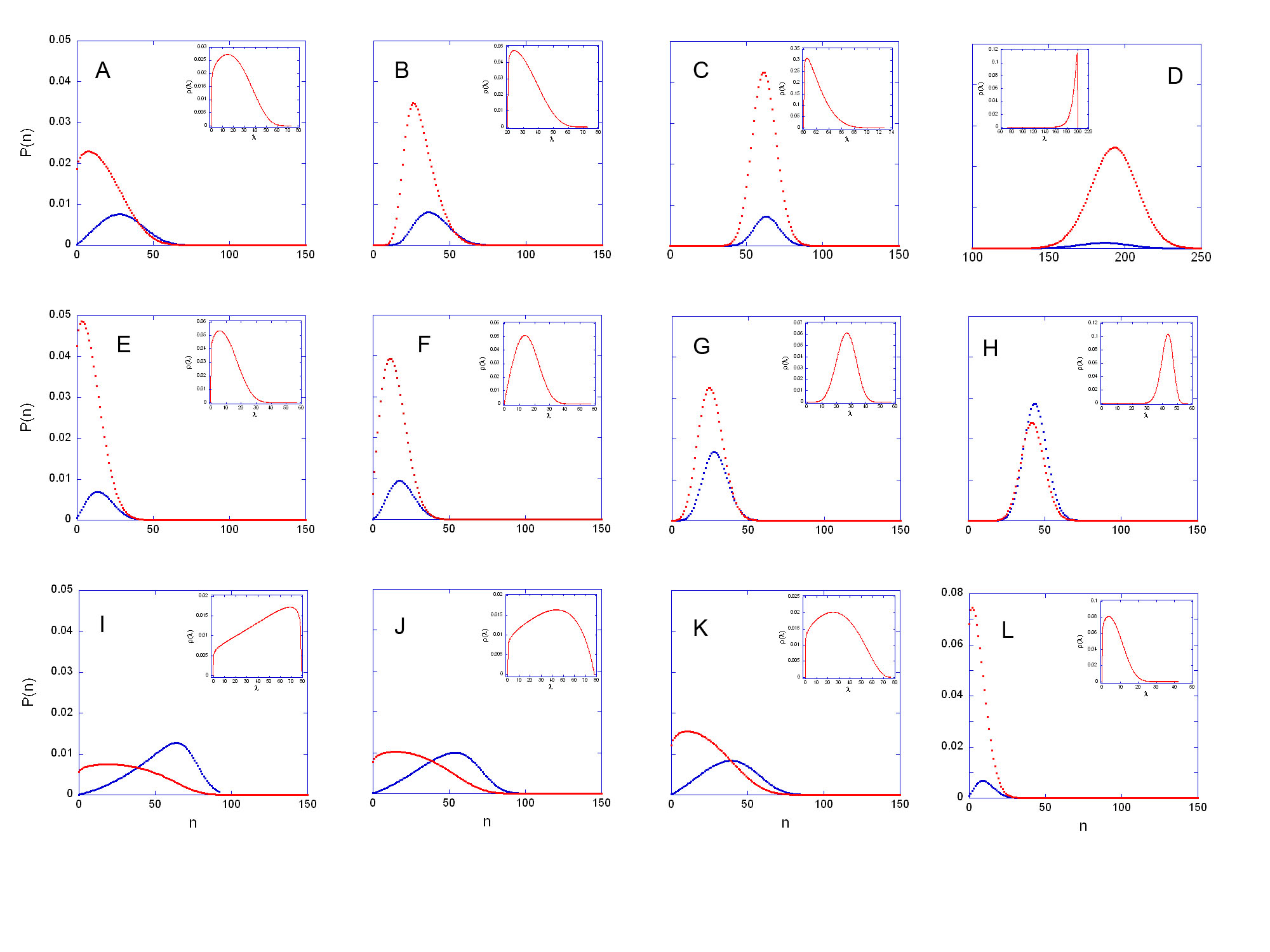}
	\caption{Calculated steady state probability distributions at twelve different parameter sets. Blue: $P(n)=P_\alpha(n,\infty)$, and red: $P(n)=P_\beta(n,\infty)$ where $n$ is the number of unbound proteins.}
\end{figure}

At each subfigure the inset shows the respective $\rho(\lambda)=\rho_\alpha(\lambda)+\rho_\beta(\lambda)$ function. The total probability of the bound state of the gene, i.e.: $P_\beta = \sum\limits_{n=0}^\infty P_\beta(n,\infty)$, and the model parameter values are given in the legends to the subfigures. 

a) $P_\beta$=0.728; $\frac{k_b}{k}$=0.9,$\frac{f}{k}$=0.2,$\frac{g_\alpha}{k}$=80,$\frac{g_\beta}{k}$=0,$\frac{h}{k}$=0.1;  

b) $P_\beta$=0.779; $\frac{k_b}{k}$=0.9,$\frac{f}{k}$=0.2,$\frac{g_\alpha}{k}$=80,$\frac{g_\beta}{k}$=20,$\frac{h}{k}$=0.1;  

c) $P_\beta$=0.853; $\frac{k_b}{k}$=0.9,$\frac{f}{k}$=0.2,$\frac{g_\alpha}{k}$=80,$\frac{g_\beta}{k}$=60,$\frac{h}{k}$=0.1;  

d) $P_\beta$=0.944; $\frac{k_b}{k}$=0.9,$\frac{f}{k}$=0.2,$\frac{g_\alpha}{k}$=80,$\frac{g_\beta}{k}$=200,$\frac{h}{k}$=0.1;  

e) $P_\beta$=0.854; $\frac{k_b}{k}$=0.9,$\frac{f}{k}$=0.2,$\frac{g_\alpha}{k}$=80,$\frac{g_\beta}{k}$=0,$\frac{h}{k}$=0.4; 

f) $P_\beta$=0.801; $\frac{k_b}{k}$=0.9,$\frac{f}{k}$=1,$\frac{g_\alpha}{k}$=80,$\frac{g_\beta}{k}$=0,$\frac{h}{k}$=0.4; 

g) $P_\beta$=0.66; $\frac{k_b}{k}$=0.9,$\frac{f}{k}$=5,$\frac{g_\alpha}{k}$=80,$\frac{g_\beta}{k}$=0,$\frac{h}{k}$=0.4; 

h) $P_\beta$=0.456; $\frac{k_b}{k}$=0.9,$\frac{f}{k}$=20,$\frac{g_\alpha}{k}$=80,$\frac{g_\beta}{k}$=0,$\frac{h}{k}$=0.4; 

i) $P_\beta$=0.435; $\frac{k_b}{k}$=0.9,$\frac{f}{k}$=0.2,$\frac{g_\alpha}{k}$=80,$\frac{g_\beta}{k}$=0,$\frac{h}{k}$=0.015; 

j) $P_\beta$=0.522; $\frac{k_b}{k}$=0.9,$\frac{f}{k}$=0.2,$\frac{g_\alpha}{k}$=80,$\frac{g_\beta}{k}$=0,$\frac{h}{k}$=0.025; 

k) $P_\beta$=0.635; $\frac{k_b}{k}$=0.9,$\frac{f}{k}$=0.2,$\frac{g_\alpha}{k}$=80,$\frac{g_\beta}{k}$=0,$\frac{h}{k}$=0.05; 

l) $P_\beta$=0.899; $\frac{k_b}{k}$=0.9,$\frac{f}{k}$=0.2,$\frac{g_\alpha}{k}$=80,$\frac{g_\beta}{k}$=0,$\frac{h}{k}$=0.9.

\newpage

\begin{figure}
	\centering
	\includegraphics[width=15cm]{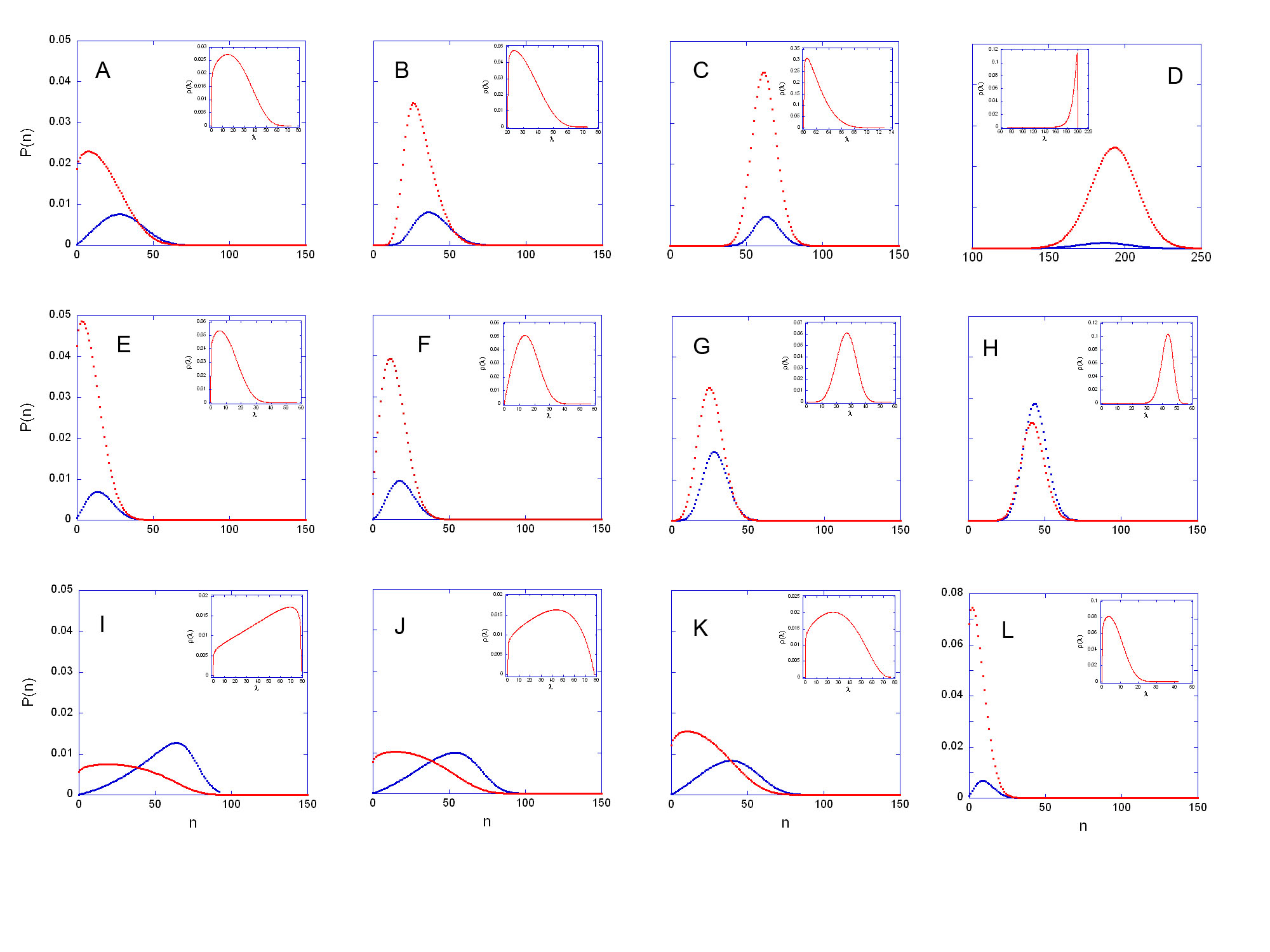}
	\caption{Calculated steady state probability distributions at twelve different parameter sets. Blue: $P(n) = P_\alpha(n,\infty)$, and red: $P(n) = P_\beta(n,\infty)$ where $n$ is the number of unbound proteins.}
\end{figure}

At each subfigure the inset shows the respective $\rho(\lambda) = \rho_\alpha(\lambda) + \rho_\beta(\lambda)$ function. The total probability of the bound state of the gene, i.e.: $P_\beta = \sum\limits_{n=0}^{\infty}P_\beta(n,\infty)$, and the model parameter values are given in the legends to the subfigures. The model parameter values in the subfigures are: 

a) $P_\beta$=0.213; $\frac{k_b}{k}$=0.66,$\frac{f}{k}$=9.3,$\frac{g_\alpha}{k}$=4.6,$\frac{g_\beta}{k}$=4.8,$\frac{h}{k}$=0.6;  

b) $P_\beta$=0.311; $\frac{k_b}{k}$=0.66,$\frac{f}{k}$=9.3,$\frac{g_\alpha}{k}$=4.6,$\frac{g_\beta}{k}$=16,$\frac{h}{k}$=0.6;  

c) $P_\beta$=0.542; $\frac{k_b}{k}$=0.66,$\frac{f}{k}$=9.3,$\frac{g_\alpha}{k}$=4.6,$\frac{g_\beta}{k}$=36,$\frac{h}{k}$=0.6;  

d) $P_\beta$=0.742; $\frac{k_b}{k}$=0.66,$\frac{f}{k}$=9.3,$\frac{g_\alpha}{k}$=4.6,$\frac{g_\beta}{k}$=66,$\frac{h}{k}$=0.6;  

e) $P_\beta$=0.592; $\frac{k_b}{k}$=0.66,$\frac{f}{k}$=1.3,$\frac{g_\alpha}{k}$=4.6,$\frac{g_\beta}{k}$=6,$\frac{h}{k}$=0.6; 

f) $P_\beta$=0.518; $\frac{k_b}{k}$=0.66,$\frac{f}{k}$=2,$\frac{g_\alpha}{k}$=4.6,$\frac{g_\beta}{k}$=6,$\frac{h}{k}$=0.6

g) $P_\beta$=0.222; $\frac{k_b}{k}$=0.66,$\frac{f}{k}$=9.3,$\frac{g_\alpha}{k}$=4.6,$\frac{g_\beta}{k}$=6,$\frac{h}{k}$=0.6

h) $P_\beta$=0.0835; $\frac{k_b}{k}$=0.66,$\frac{f}{k}$=30,$\frac{g_\alpha}{k}$=4.6,$\frac{g_\beta}{k}$=6,$\frac{h}{k}$=0.6

i) $P_\beta$=0.0232; $\frac{k_b}{k}$=0.66,$\frac{f}{k}$=9.3,$\frac{g_\alpha}{k}$=4.6,$\frac{g_\beta}{k}$=6,$\frac{h}{k}$=0.05

j) $P_\beta$=0.0856; $\frac{k_b}{k}$=0.66,$\frac{f}{k}$=9.3,$\frac{g_\alpha}{k}$=4.6,$\frac{g_\beta}{k}$=6,$\frac{h}{k}$=0.2

k) $P_\beta$=0.276; $\frac{k_b}{k}$=0.66,$\frac{f}{k}$=9.3,$\frac{g_\alpha}{k}$=4.6,$\frac{g_\beta}{k}$=6,$\frac{h}{k}$=0.8

l) $P_\beta$=0.749; $\frac{k_b}{k}$=0.66,$\frac{f}{k}$=9.3,$\frac{g_\alpha}{k}$=4.6,$\frac{g_\beta}{k}$=6,$\frac{h}{k}$=6

\newpage
\section*{References}
\doublespacing

[1] L. H. Hartwell\textit{ et al.}, \textit{Nature} \textbf{402}, C47 (1999)

[2] S. S. Shen-Orr\textit{ et al.}, \textit{Nature Genetics} \textbf{31}, 64 (2002)

[3] T. I. Lee\textit{ et al.}, \textit{Science} \textbf{298}, 799 (2002)

[4] A. Ma'ayan\textit{ et al.}, \textit{Science} \textbf{309}, 1078 (2005)

[5] D. T. Gillespie, \textit{Journal of Physical Chemistry} \textbf{81}, 2340 (1977)

[6] F. J. Isaacs\textit{ et al.}, \textit{Proceedings of the National Academy of Sciences of the United States of America} \textbf{100}, 7714 (2003)

[7] D. Thieffry\textit{ et al.}, \textit{Bioessays} \textbf{20}, 433 (1998)

[8] J. E. M. Hornos\textit{ et al.}, \textit{Physical Review E} \textbf{72}, 051907 (2005)

[9] S. Iyer-Biswas, and C. Jayaprakash, \textit{arXiv} preprint arXiv:1110.2804  (2011)

[10] C. W. Gardiner, \textit{Stochastic Methods: A Handbook for the Natural and Social Sciences }(Springer, Berlin, 2004), Series in Synergetics.

[11] C. W. Gardiner, and S. Chaturvedi, \textit{Journal of Statistical Physics} \textbf{17}, 429 (1977)

[12] R. Grima, D. R. Schmidt, and T. J. Newman, \textit{Journal of Chemical Physics} \textbf{137},  035104 (2012)

[13] G. C. P. Innocentini, and J. E. M. Hornos, \textit{Journal of Mathematical Biology} \textbf{55}, 413 (2007)

[14] A. F. Ramos, and J. E. M. Hornos, \textit{Physical Review Letters} \textbf{99}, 108103 (2007)

[15] A. F. Ramos, G. C. P. Innocentini, and J. E. M. Hornos, \textit{Physical Review E} \textbf{83}, 062902 (2011)

[16] A. F. Ramos\textit{ et al.}, \textit{Iet Systems Biology} \textbf{4}, 311 (2010)

[17] W. H. Press\textit{ et al.}, \textit{Numerical Recipes} (Cambridge Press, New York, 1989).

\end{document}


\beginsupplement

\title{\textbf{SUPPLEMENTAL MATERIAL TO}\\Self-regulating genes. Exact steady state solution by using Poisson Representation.}
\author[1]{Istv\'an P. Sug\'ar}
\author[2]{Istv\'an Simon}
\affil[1]{\small Department of Neurology and Center for Translational Systems Biology, Ichan School of Medicine at Mount Sinai, New York, NY 10029}
\affil[2]{\small Institute of Enzymology, Research Center for Natural Sciences, Hungarian Academy of Sciences, Budapest, Hungary}
\date{\the\year}
\maketitle
{e-mail: istvan.sugar@mssm.edu, simon@enzim.hu}

\section{Comparison of the master equations}

Here we compare the master equations in the main text, Eqs.7-8, with the master equations in the papers of Hornos and his coworkers [1-5]. Our self-regulating gene model, given by Eqs.1-6, is equivalent with the models discussed in the above papers when the degradation rate of the bound protein is zero, i.e. $\frac{k_b}{k}=0$. In spite of the similarity of the model, the master equations in the above papers are different from Eqs.7-8.

By using the notations of Eqs.7-8 the master equations in the papers of Ramos et al. [3-5] are: 

\begin{equation}
\frac{dP_\alpha(n,t)}{dt}= f P_\beta(n,t)+g_\alpha P_\alpha(n-1,t) + k[n+1] P_\alpha(n+1,t) - [hn+g_\alpha+kn]P_\alpha(n,t)
\end{equation}

\begin{equation}
\frac{dP_\beta(n,t)}{dt}= k [n+1]P_\beta(n+1,t)+g_\beta P_\beta(n-1,t)+hnP_\alpha(n,t)-[f+g_\beta+nk]P_\beta(n,t)
\end{equation}

It is important to mention that in Ramos’s equations, Eqs.S1-S2, $n$ is defined as the number of ‘free’ proteins, i.e. unbound proteins.

Note that the 1\textsuperscript{st} term on the right hand side and the 3\textsuperscript{rd} term on the right hand side of Ramos’s first and second equations, respectively are different from the respective terms in Eqs.7-8. In Ramos’s first equation (Eq.S1) it is not taken into consideration that the protein dissociation process increases the number of unbound proteins by 1, while in Ramos’s second equation (Eq.S2) it is not taken into consideration that the protein binding process decreases the number of unbound proteins by 1. Ramos et al. [4] obtained the exact time dependent solutions of Eqs.S1,S2 by using the Heun functions.

However, because of the above mentioned differences between our and Ramos’s master equations we are unable to get the exact time dependent solutions of our master equations (Eqs.7-8). 

The master equations in Hornos’s paper [1] look similar to Ramos’s equations[3-5] but \textit{n} is defined as the total number of proteins (bound and unbound). For clarity let $\overline n$ be the total number of proteins. Thus, by using our notations (in Eqs.7-8), Hornos’s equations are:

\begin{equation}
\frac{dP_\alpha(\overline{n},t)}{dt} = f P_\beta(\overline{n},t)+g_\alpha P_\alpha(\overline{n}-1,t) + k[\overline{n}+1] P_\alpha(\overline{n}+1,t)-[h\overline{n}+g_\alpha+k\overline{n}]P_\alpha(\overline{n},t)
\end{equation}

\begin{equation}
\frac{dP_\beta(\overline{n},t)}{dt} = k[\overline{n}+1]P_\beta(\overline{n}+1,t)+g_\beta P_\beta(\overline{n}-1,t)+h\overline{n}P_\alpha(\overline{n},t)-[f+g_\beta+\overline{n}k]P_\beta(\overline{n},t)
\end{equation}

In order to compare Hornos’s equations with our master equations, let us rewrite Eqs.7-8 by using $\overline{n}$ instead of $n$:

\begin{equation}
\frac{dP_\alpha(\overline{n},t)}{dt}=fP_\beta(\overline{n},t)+k_bP_\beta(\overline{n}+1,t)+g_\alpha P_\alpha(\overline{n}-1,t)+k[\overline{n}+1] P_\alpha(\overline{n}+1,t)-[h\overline{n}+g_\alpha+k\overline{n}] P_\alpha(\overline{n},t)
\end{equation}

\begin{equation}
\frac{dP_\beta(\overline{n},t)}{dt} = k\overline{n}P_\beta(\overline{n}+1,t)+g_\beta P_\beta(\overline{n}-1,t)+h\overline{n}P_\alpha(\overline{n},t)-[f+g_\beta+k(\overline{n}-1)+k_b]P_\beta(\overline{n},t)
\end{equation}

Let us compare now Hornos’s equations (Eqs.S3-S4) with our rewritten master equations (Eqs.S5-S6) when $\frac{k_b}{k} = 0$. Hornos’s first equation (Eq.S3) agrees with our first equation (Eq.S5). However, in the second set of equations (Eq.S4 and Eq.S6) there are two differences: one in the 1\textsuperscript{st} term and one in the last term of the right hand side. This is because Hornos does not take into consideration that in the $\beta$ state out of $\overline{n}$ proteins, only the unbound proteins, i.e. $\overline{n}-1$ proteins, can degrade.

Finally, let us compare Hornos’s equations (Eqs.S3-S4) with our rewritten master equations (Eqs.S5-S6), when $\frac{k_b}{k}=1$, i.e. the degradation rate of the bound and unbound proteins are similar. In this case our rewritten master equations are:

\begin{equation}
\frac{dP_\alpha(\overline{n},t)}{dt}= f P_\beta(\overline{n},t)+  k P_\beta(\overline{n}+1,t)+g_\alpha P_\alpha(\overline{n}-1,t) + k[\overline{n}+1]P_\alpha(\overline{n}+1,t)-[h\overline{n}+g_\alpha+k\overline{n}]P_\alpha(\overline{n},t)
\end{equation}

\begin{equation}
\frac{dP_\beta(\overline{n},t)}{dt}= k \overline{n}P_\beta(\overline{n}+1,t)+g_\beta P_\beta(\overline{n}-1,t)+h\overline{n}P_\alpha(\overline{n},t)-[f+g_\beta+k\overline{n}]P_\beta(\overline{n},t)
\end{equation}

In Eq.S7 the 2\textsuperscript{nd} term on the right hand side is missing in Hornos’s 1\textsuperscript{st} equation (Eq.S3), while in Eq.S8 the 1\textsuperscript{st} term on the right hand side is different from the respective term in Hornos’s equation (Eq.S4).

\section{Solving the master equations by using Poisson Representation}

Let us solve Eqs.7,8 at steady state by using the method of Poisson representation [6].  

We assume the existence of $\rho_\alpha(\lambda)$ and $\rho_\beta(\lambda)$ functions that yield $P_\alpha(n,\infty)$ and $P_\beta(n,\infty)$ by

\begin{equation}
P_i(n,\infty)=\int\limits_A^B d\lambda \rho_i(\lambda)e^{-\lambda} \frac{\lambda^n}{n!}
\end{equation}

where $i=\alpha$ or $\beta$.

After substituting the above forms of the probability functions into the generating functions in Eqs.11,12 we get the following equations for the $\rho_\alpha(\lambda)$ and $\rho_\beta(\lambda)$ functions

\begin{eqnarray}
0 & = &\bigg[ f \rho_\beta(B)+ \Big\{ g_\alpha-hB-kB\Big\} \rho_\alpha(B) \bigg] e^{B(s-1)} \notag \\
& & \hspace{-0.8em}-\bigg[f\rho_\beta(A)+\Big\{g_\alpha-hA-kA\Big\}\rho_\alpha(A)\bigg] e^{A(s-1)} \notag \\
& & \hspace{-0.8em}+\int\limits_A^B d\lambda \bigg[ f \rho_\beta-f \frac{d\rho_\beta}{d\lambda}+k_b \rho_\beta-g_\alpha\frac{d\rho_\alpha}{d\lambda}+k \frac{d(\lambda\rho_\alpha)}{d\lambda}+h\frac{d(\lambda\rho_\alpha)}{d\lambda}-h\lambda\rho_\alpha \bigg] e^{\lambda(s-1)}\\ \notag \\
0 & = &\Big[g_\beta-kB\Big]\rho_\beta(B)e^{B(s-1)} - \Big[g_\beta-kA\Big]\rho_\beta(A)e^{A(s-1)} \notag \\
& & +\int\limits_A^B d\lambda k \bigg[ \frac{d(\lambda\rho_\beta)}{d\lambda}-g_\beta \frac{d\rho_\beta}{d\lambda}+h\lambda\rho_\alpha-f\rho_\beta-k_b \rho_\beta \bigg] e^{\lambda(s-1)}
\end{eqnarray}

Deriving Eqs.S10,S11 the following relationships were utilized:

\begin{eqnarray*}
G_i & \equiv & G_i(s,\infty) = \sum\limits_{n=0}^\infty s^n \int\limits_A^B d\lambda \rho_i(\lambda)e^{-\lambda} \frac{\lambda^n}{n!} = \int\limits_A^B d\lambda\rho_i(\lambda)e^{-\lambda} \left[ \sum\limits_{n=0}^\infty \frac{(s\lambda)^n}{n!} \right] = \int\limits_A^B d\lambda\rho_i(\lambda)e^{\lambda(s-1)}
\end{eqnarray*}

\begin{eqnarray*}
sG_i & = & (s-1)\int\limits_A^B d\lambda \rho_i e^{\lambda(s-1)}+G_i \\
 & = & -\int\limits_A^B d\lambda \frac{d\rho_i}{d\lambda} e^{\lambda(s-1)}+ \int\limits_A^B d\lambda \rho_i e^{\lambda(s-1)} +\rho_i(B)e^{B(s-1)}- \rho_i(A)e^{A(s-1)} \\ \notag \\
s\frac{\partial G_i}{\partial s} & = & (s-1) \int\limits_A^B d\lambda(\rho_i\lambda) e^{\lambda(s-1)} + \int\limits_A^B d\lambda(\rho_i\lambda)e^{\lambda(s-1)} \\ \notag \\
 &  = & -\int\limits_A^B d\lambda \frac{d(\lambda\rho_i)}{d\lambda} e^{\lambda(s-1)} + \int\limits_A^B d\lambda(\rho_i\lambda)e^{\lambda(s-1)} +\rho_i(B) B e^{B(s-1)}-\rho_i(A)Ae^{A(s-1)} \\\notag \\
\frac{\partial G_i}{\partial s} & = & \int\limits_A^B d\lambda(\rho_i\lambda)e^{\lambda(s-1)}
\end{eqnarray*}

where in the 2\textsuperscript{nd} and 3\textsuperscript{rd} equations we integrated by parts.

\subsection{Eliminating terms at integration boundaries}

\textbf{\textit{}}

In Eqs.S10,S11 the first two terms can be eliminated by properly choosing the integration boundaries $A$ and $B$. That is

\begin{eqnarray*}
0 & = & \bigg[f\rho_\beta(B) + \Big\{ g_\alpha-hB-kB \Big\} \rho_\alpha(B)\bigg]e^{B(s-1)}
 -\bigg[f\rho_\beta(A) + \Big\{ g_\alpha-hA-kA \Big\} \rho_\alpha(A)\bigg]e^{A(s-1)} \\ \notag \\
0 & = & \Big[ g_\beta-kB \Big] \rho_\beta(B)e^{B(s-1)}
  -\Big[ g_\beta-kA \Big] \rho_\beta(A)e^{A(s-1)}
\end{eqnarray*}

if

\begin{equation}
\rho_\alpha(A) = \rho_\beta(A) = \rho_\alpha(B) = \rho_\beta(B) = 0
\end{equation}

\subsection{Determining $\rho_\alpha(\lambda)$ and $\rho_\beta(\lambda)$ functions}

By determining the $\rho_\alpha(\lambda)$ and $\rho_\beta(\lambda)$ functions one can get the values of $A$ and $B$ where Eqs.S12 are valid. In order to calculate these functions let us take the sum of Eq.S10 and Eq.S11:

\begin{equation}
0=\int\limits_A^B d\lambda \bigg\{\frac{d}{d\lambda} \Big[ (k\lambda-f-g_\beta) \rho_\beta(\lambda)
+ (k\lambda+h\lambda-g_\alpha) \rho_\alpha(\lambda) \Big] \bigg\} e^{\lambda(s-1)}
\end{equation}

From Eq.S13 a relationship between $\rho_\alpha(\lambda)$ and $\rho_\beta(\lambda)$ follows

\begin{equation}
\Big[ k\lambda-f-g_\beta \Big] \rho_\beta(\lambda)+
\Big[ k\lambda+h\lambda-g_\alpha \Big] \rho_\alpha(\lambda) = C
\end{equation}

where, as a consequence of Eqs.S12, the constant, $C$, is equal to zero.

By properly choosing the integration boundaries the first two terms becomes zero in Eq.S11 and we get the following differential equation: 

\begin{equation}
k \frac{d[ \lambda \rho_\beta(\lambda)]}{d\lambda}-
g_\beta \frac{d \rho_\beta(\lambda)}{d\lambda} + h\lambda\rho_\alpha(\lambda)-
f\rho_\beta(\lambda)-k_b\rho_\beta(\lambda) = 0
\end{equation}

After substituting Eq.S14 into Eq.S15 we get:

\begin{equation}
\frac{d\rho_\beta}{\rho_\beta} = \left[ -\frac{k-k_b-f}{k \left(\lambda-\frac{g_\beta}{k}\right)} +
\frac{h}{k(k+h)} \frac{\lambda(k\lambda-f-g_\beta)}{(\lambda-\frac{g_\beta}{k})\left(\lambda-\frac{g_\alpha}{k+h} \right)} \right] d\lambda = 
\left[ E+\frac{F-L}{\lambda-\frac{g_\beta}{k}}+\frac{G}{\lambda-\frac{g_\alpha}{k+h}} \right] d\lambda
\end{equation}

where 

\begin{eqnarray*}
E & = & \frac{h}{(k+h)}\\
F & = & \frac{fhg_\beta}{ k^2 (k+h) \left( \frac{g_\alpha}{k+h}-\frac{g_\beta}{k} \right) }\\
G & = & \frac{h}{k+h} \left[ \frac{g_\alpha}{k+h}-\frac{f+g_\beta}{k} \right] \frac{\frac{g_\alpha}{k+h}}{\frac{g_\alpha}{k+h} - \frac{g_\beta}{k}}\\
L & = & \frac{(k-k_b-f)}{k}
\end{eqnarray*}

After integrating Eq.S16 we get the steady-state solution for $\rho_\beta$:

\begin{equation}
\rho_\beta(\lambda) = D \left| \lambda-\frac{g_\beta}{k} \right|^{F-L} \left| \lambda-\frac{g_\alpha}{k+h}\right|^G e^{E\lambda}
\end{equation}

where $D$ is the integration constant. By substituting Eq.S17 into Eq.S14 we get the steady-state solution for 
$\rho_\alpha$:

\begin{eqnarray}
\rho_\alpha(\lambda) & = & D \bigg[ \frac{f+g_\beta-k\lambda}{k\lambda+h\lambda-g_\alpha} \bigg] \left| \lambda-\frac{g_\beta}{k} \right|^{F-L} \left| \lambda-\frac{g_\alpha}{k+h}\right|^G e^{E\lambda} \notag \\
& = & \left[ \frac{D}{k+h} \right] \left[ \frac{f+g_\beta-k\lambda}{\lambda-\frac{g_\alpha}{k+h}} \right] \left| \lambda-\frac{g_\beta}{k} \right|^{F-L} \left| \lambda-\frac{g_\alpha}{k+h}\right|^G e^{E\lambda}
\end{eqnarray}

\subsection{Determining integration boundaries}

From Eqs.S17,S18 one can get the integration boundaries $A$ and $B$ where Eqs.S12 are fulfilled. These integration boundaries are listed in the main text (Table 1).

\section{Normalization condition}

By means of the following normalization condition we can calculate the integration constant, $D$:

\begin{eqnarray}
1 & = & \sum\limits_{n=0}^\infty P_\alpha(n,\infty) + \sum\limits_{n=0}^\infty P_\beta(n,\infty) \notag \\
  & = &\sum\limits_{n=0}^\infty \int_A^B d\lambda \rho_\alpha(\lambda)e^{-\lambda} \frac{\lambda^n}{n!} +
\sum\limits_{n=0}^\infty \int_A^B d\lambda \rho_\beta(\lambda) e^{-\lambda} \frac{\lambda^n}{n!} \notag \\
& = & \int\limits_A^B d\lambda \rho_\alpha(\lambda) \left[ \sum\limits_{n=0}^\infty e^{-\lambda} \frac{\lambda^n}{n!} \right] + 
\int\limits_A^B d\lambda \rho_\beta(\lambda) \left[ \sum\limits_{n=0}^\infty e^{-\lambda} \frac{\lambda^n}{n!} \right] \notag \\
& = & \int\limits_A^B d\lambda[ \rho_\alpha(\lambda) + \rho_\beta(\lambda)] \notag \\
& = & D \int\limits_A^B d\lambda \left|\lambda-\frac{g_\beta}{k}\right|^{F-L} \left|\lambda-\frac{g_\alpha}{k+h}\right|^G e^{E\lambda} \left[ 1-\frac{k\lambda-f-g_\beta}{k\lambda+h\lambda-g_\alpha} \right]
\end{eqnarray}

\section{Closed form solutions}

When either $g_\alpha k>g_\beta k=0$ or $g_\beta k>g_\alpha k=0$ one can get $P_\alpha(n,\infty)$ and $P_\beta(n,\infty)$ in closed form.

According to the integral representations of the confluent hypergeometric function[7]

\begin{equation}
\int\limits_0^b d\lambda \lambda^{\mu-1}(b-\lambda)^{\nu-1} e^{\beta\lambda} = 
\frac{\Gamma(\mu)\Gamma(\nu)}{\Gamma(\mu+\nu)} b^{\mu+\nu-1} M(\mu,\mu+\nu,b\beta)
\end{equation}

where $M$  is the Kummer’s confluent hypergeometric function, and Eq.S20 holds when $\mu>0$ and $u>0$.
When $g_\alpha k>g_\beta k=0$ the exact steady-state solutions are:

\begin{eqnarray}
P_\beta(n,\infty) & = & \frac{D}{n!} \int\limits_0^b d\lambda \lambda^{-L} (b-\lambda)^G e^{E\lambda} \lambda^n e^{-\lambda} \notag \\
& = & Db^{G+n+1-L} \frac{\Gamma(G+1)\Gamma(n+1-L)}{\Gamma(n+1)\Gamma(G+n+2-L)} M(1+n-L,2+n+G-L,Eb-b)
\end{eqnarray}

\begin{eqnarray}
P_\alpha(n,\infty) & = & \frac{D}{n!} \int\limits_0^b d\lambda \lambda^n e^{-\lambda} e^{E\lambda} \left[ -\frac{f}{k+h} \lambda^{-L} (b-\lambda)^{G-1} + \frac{k}{k+h} \lambda^{-L+1} (b-\lambda)^{G-1} \right] \notag \\
& = & -\frac{Df}{k+h}b^{G+n-L}  \frac{\Gamma(n+1-L)\Gamma(G)}{\Gamma(n+1)\Gamma(G+n+1-L)} M(1+n-L,1+n+G-L,Eb-b) \notag \\
& & +\frac{Dk}{k+h}b^{G+n+1-L}\frac{\Gamma(n+2-L)\Gamma(G)}{\Gamma(n+1)\Gamma(G+n+2-L)} M(2+n-L,2+n+G-L,Eb-b) \notag \\
& & 
\end{eqnarray}

where $b=\frac{g_\alpha}{k+h}$. 

When $g_\beta k>g_\alpha k=0$ the exact steady-state solutions are:

\begin{eqnarray}
P_\beta(n,\infty) & = & \frac{D}{n!} \int\limits_0^b d\lambda \lambda^G (b-\lambda)^{F-L}e^{E\lambda}\lambda^n e^{-\lambda} \notag \\
& = & Db^{G+n+1+F-L} \frac{\Gamma(G+n+1)\Gamma(1+F-L)}{\Gamma(n+1)\Gamma(G+n+2+F-L)} M(1+G+n,2+n+G+F-L,Eb-b) \notag \\
& &
\end{eqnarray}

\begin{eqnarray}
P_\alpha(n,\infty) & = & \frac{D}{(k+h)n!} \int\limits_0^b d\lambda \lambda^n e^{-\lambda} e^{E\lambda} \bigg[(f+g_\beta-k\lambda)\lambda^{G-1} (b-\lambda)^{F-L} \bigg] \notag \\
& = & \frac{D(f+g_\beta)}{k+h}b^{G+n+F-L} \frac{\Gamma(1+F-L)\Gamma(G+n)}{\Gamma(n+1)\Gamma(G+n+1+F-L)} M(G+n,1+n+G+F-L,Eb-b) \notag \\
& & - \frac{Dk}{k+h}b^{G+n+1+F-L} \frac{\Gamma(1+F-L)\Gamma(G+n+1)}{\Gamma(n+1)\Gamma(G+n+2+F-L)} M(1+G+n,2+n+G+F-L,Eb-b) \notag \\
& & 
\end{eqnarray}

where $b=\frac{g_\beta}{k}$.

\section{Consistency check between the solution and the master equations}

In the case of steady state and at $n=0$ the master equations (Eqs.7,8) are:

\begin{eqnarray}
0 & = & k_bP_\beta(0,\infty) + kP_\alpha(1,\infty)-g_\alpha P_\alpha(0,\infty) \\ \notag \\
0 & = & kP_\beta(1,\infty)+hP_\alpha(1,\infty)-\Big[f+k_b+g_\beta\Big]P_\beta(0,\infty)
\end{eqnarray}

After eliminating $P_\alpha(1,\infty)$ from these equations we get

\begin{equation}
\frac{P_\alpha(0,\infty)}{P_\beta(0,\infty)} = \frac{k_b}{g_\alpha} + \frac{kf}{hg_\alpha} + \frac{kk_b}{hg_\alpha} + \frac{kg_\beta}{h g_\alpha} - \frac{k^2}{hg_\alpha} \frac{P_\beta(1,\infty)}{P_\beta(0,\infty)}
\end{equation}

The master equations are consistent with the solution (given by Eq.13) if the above equality holds after substituting the values of $P_\alpha(0,\infty)$, $P_\beta(0,\infty)$ and $P_\beta(1,\infty)$. At the following parameter set:  $\frac{h}{k}=0.1$, $\frac{k_b}{k}=0.9$, $\frac{f}{k}=0.2$, $\frac{g_\alpha}{k}=80$, $\frac{g_\beta}{k}=0$ the solution provides: $P_\alpha(0,\infty)=0.00021673$, $P_\beta(0,\infty)=0.018694$ and $P_\beta(1,\infty)=0.020516$. After substituting the above parameter values and solutions into Eq.S27 the numerical value of the left hand side, 0.011593559, agrees with that of the right hand side, 0.011566947, up to 4 decimals.

\newpage
\section*{References}
\doublespacing

[1] J. E. M. Hornos\textit{ et al.}, \textit{Physical Review E} \textbf{72, }051907 (2005)

[2] G. C. P. Innocentini, and J. E. M. Hornos, \textit{Journal of Mathematical Biology} \textbf{55, }413 (2007)

[3] A. F. Ramos, and J. E. M. Hornos, \textit{Physical Review Letters} \textbf{99, }108103 (2007)

[4] A. F. Ramos, G. C. P. Innocentini, and J. E. M. Hornos, \textit{Physical Review E} \textbf{83, }062902 (2011)

[5] A. F. Ramos\textit{ et al.}, \textit{Iet Systems Biology} \textbf{4, }311 (2010)

[6] C. W. Gardiner, \textit{Stochastic Methods: A Handbook for the Natural and Social Sciences }(Springer, Berlin, 2004), Series in Synergetics.

[7] M. Abramowitz, and I. A. Sregun, \textit{Handbook of Mathematical Functions }(National Bureau of Standards, 1972).